\documentclass{pasj01}
\Received{$\langle$reception date$\rangle$}
\Accepted{$\langle$acception date$\rangle$}
\Published{$\langle$publication date$\rangle$}

\usepackage{ulem} 
\usepackage{graphicx}
\graphicspath{{.\\}}
\begin{document}

\title{Evidence for GeV Cosmic Rays from White Dwarfs in the Local Cosmic Ray
Spectra and in the Gamma-ray Emissivity of the Inner Galaxy}
\author{Tuneyoshi KAMAE\altaffilmark{1,2,*}}%
\author{Shiu-Hang LEE\altaffilmark{3}}%
\author{Kazuo MAKISHIMA\altaffilmark{4}}%
\author{Shinpei SHIBATA\altaffilmark{5}}%
\author{Toshikazu SHIGEYAMA\altaffilmark{6}}%
\altaffiltext{1}{Department of Physics, Graduate School of Science, The University of Tokyo, 7-3-1
Hongo, Bunkyo-ku, Tokyo 113-0033, Japan}
\altaffiltext{2}{SLAC National Laboratory and The Kavli Institute for Particle Astrophysics and
Cosmology, Stanford University, 2575 Sand Hill Road, Menlo Park, CA 94025 USA}
\altaffiltext{3}{Department of Astronomy, Kyoto University, Kitashirakawa, Oiwake-cho, Sakyo-ku, Kyoto
606-8502, Japan}
\altaffiltext{4}{Makishima Cosmic Radiation Laboratory, RIKEN, 2-1 Hirosawa, Wako, Saitama
351-0198, Japan}
\altaffiltext{5}{Department of Physics, Graduate School of Science, Yamagata University, 1-4-12,
Kojirakawa-machi, Yamagata City, Yamagata 990-8560, Japan}
\altaffiltext{6}{Research Center for the Early Universe, Graduate School of Science, The University of
Tokyo, 7-3-1 Hongo, Bunkyo-ku, Tokyo 113-0033, Japan}

\email{kamae@slac.stanford.edu}

\maketitle

\begin{abstract}
Recent observations found that electrons are accelerated to $\sim$10 GeV and emit synchrotron hard X-rays in two magnetic white dwarfs (WDs), also known as cataclysmic variables (CVs). In nova outbursts of WDs, multi-GeV gamma-rays were detected inferring that protons are accelerated to 100 GeV or higher. In recent optical surveys, the WD density is found to be higher near the Sun than in the Galactic disk by a factor $\sim$2.5. The cosmic rays (CR) produced by local CVs and novae will accumulate in the local bubble for $10^6$ -- $10^7$ yrs. On these findings, we search for CRs from historic CVs and novae in the observed CR spectra. We model the CR spectra at the heliopause as sums of Galactic and local components based on observational data as much as possible. The initial Galactic CR electron and proton spectra are deduced from the gamma-ray emissivity, the local electron spectrum from the hard X-ray spectra at the CVs, and the local proton spectrum inferred by gamma-ray spectrum at novae. These spectral shapes are then expressed in a simple set of polynomial functions of CR energy and regressively fitted until the high-energy ($>$100 GeV) CR spectra near Earth and the Voyager-1 spectra at the heliopause are reproduced. We then extend the modeling to nuclear CR spectra and find that one spectral shape fits all local nuclear CRs and the apparent hardening of the nuclear CR spectra is caused by the roll-down of local nuclear spectra around 100 -- 200 GeV. All local CR spectra populate in a limited energy band below 100 -- 200 GeV and enhance gamma-ray emissivity below $\sim$10 GeV. Such an enhancement is observed in the inner Galaxy, suggesting the CR fluxes from CVs and novae are substantially higher there.   
\end{abstract}

\section{Introduction}

White dwarfs (WD) are the final evolutionary state for stars with mass below $\sim$8 solar mass and constitute about 97\% of stars in the Milky Way (Fontaine et al.\ 2001). They are grouped into ``isolated or single'', ``binary'', ``Sirius-like'', ``double degenerate'' and ``magnetic'' (Sion et al.\ 2009; Giammichele, Bergeron, \& Dufour\ 2012; Holberg et al.\ 2016). Isolated WDs dominate over binary WDs in number and are classified by their spectral type or the dominant atomic composition near the surface as: DA (H), DB (He), DZ (He and metals), and DQ (carbon). About 10\% of local WDs are now known to be in the binary configuration with a main sequence star and about 9\% of local WDs are magnetic WDs (Holberg et al.\ 2016). It is generally thought that a substantial fraction ($\sim$10\%) of magnetic WDs are now or were in binary configuration. These numbers may change when new WD surveys extend the sky coverage in the future. For example, a new survey focused on the DZ type by Hollands, Gaensicke, and Koester (2015) found 79 cool DZ WDs within 13 pc of the Sun, of which seven are highly magnetized. 

Tremblay et al.\ (2014) derived, from the WD survey data by Sion et al.\ (2009), that the WD density within 20 pc of the Sun is higher by a factor $\sim$2.5 compared with that in the Galactic thin disk. The survey within $\sim$20 pc is now considered to be complete to $\sim$100\% and the local WD density comes out to be consistent with an estimate of the total WD mass of 37 solar masses within 13~pc (Reid, Gizis, \& Hawley 2002) for the average WD mass of 0.5 solar mass.

When WDs, magnetic or non-magnetic, have a companion star and accrete matter,
various phenomena occur including CVs, nova outbursts and Type Ia supernovae. WDs remain quiescent most of time and the X-ray and gamma-ray fluxes from them do not exceed the detection threshold of satellite-borne instruments. When they become active as CVs, novae, or supernovae, CRs are emitted and accumulated in the local bubble of radius around $\sim$50~pc with the Sun near the center. These CRs stay in the bubble for 10$^6$ to 10$^7$ years (Strong, Moskalenko, \& Ptuskin 2007; Erlykin, Makavariani, \& Wolfendale 2017). 

It is known that the star formation rate had been enhanced near the Sun in the past $\sim$5 Gyr (Tremblay et al.\ 2014). By now, a large fraction of these stars have turned to WDs in the local bubble, which accounts for the about 2.5 times higher density of WDs around us.

Strongly magnetized WDs in the binary configuration have magnetic field around 10$^5$ to 10$^7$ G in the pole regions, and rotate at a period around 1 hour (Terada\ 2013; Ferrario et al.\ 2015). They can induce an electric potential as high as 10$^{14}$ -- 10$^{16}$~V and accelerate particles to high energy. There has been no observational evidence, however, until the detection of synchrotron hard X-rays from AE Aquarii with Suzaku by Terada et al.\ (2008). The ratio of the observed hard X-ray luminosity to the spin-down luminosity of the WD in AE Aquarii is estimated to be 0.01 -- 0.1\%, similarly to young rotation-powered pulsars in the same energy range (Terada et al.\ 2008). Hints of electron acceleration in WDs were also found in the hard X-ray spectrum from classical nova V2491 Cyg (Takei et al.\ 2009) and in the radio synchrotron emission from recurrent nova RS Oph (O’Brien et al.\ 2006; Rupen, Mioduszewski, \& Sokoloski 2008). Marsh et al.\ (2016) detected, recently, spin-powered pulsed emission from a WD binary, AR Scorpii, whose wide band spectrum is indicative of synchrotron emission. We refer to Revnivtsev et al.\ (2008) for listing of magnetic CV binaries detected in the hard X-ray band.

Another important evidence for particle acceleration in WDs was obtained with Fermi Large Area Telescope (LAT) at classical nova in the symbiotic binary V407 Cygnii 2010 (Abdo et al.\ 2010). The LAT soon added 3 more pieces of evidence for gamma-ray novae, V959 Monocerotis 2012, V1324 Scorpii 2012, and V339 Delphini 2013 (Ackermann et al.\ 2014), establishing novae as a distinct class of gamma-ray sources, or equivalently, a class of CR acceleration sites. Their surface composition is likely to contain He, C, N, and O but observational evidence is obtained only for V407 Cygnii. Two more novae were detected recently at lower statistical significance, V1369 Centauri 2013 and V5668 Sagittarii 2015 (Cheung et al.\ 2016). 

The complexity of the WD binary system and the nova outburst process makes theoretical study on the acceleration mechanism difficult. Geng, Zhang, and Huang (2016) considered two possible electron acceleration sites in the AE Scorpii system, one on the plane where the pressure from the WD magnetosphere balances against the wind pressure from the companion star. The other possible site is located where the open magnetic field lines from the WD touch the pressure-balance surface.

Metzger et al.\ (2015) analyzed gamma-ray emitting novae detected with Fermi LAT and disfavored the leptonic scenario because the implied acceleration efficiency is much higher than that expected for electrons. We note that Vurm and Metzger (2016) discussed about the hard X-ray and gamma-ray emission in non-relativistic radiative shocks, the most likely acceleration environment in the gamma-ray nova system.

On the direct CR spectral measurement side, the precision has reached a new level in the last decade (PAMELA: Adriani et al.\ 2011a, 2011b, 2013; AMS-02: Aguilar et al. 2014, 2015a, 2015b, Yan et al.\ 2017, Aguilar et al.\ 2017). The electron, proton, He, C, N, and O spectra measured at the heliopause by Voyager-1 (Stone et al.\ 2013, Cummings et al.\ 2016) are particularly important in probing CR spectra free from influence of the solar effects. Spectral hardening in nuclear CRs is an important feature these measurements uncovered together with earlier measurements (ATIC-2: Wefel et al.\ 2008; CREAM: Ahn et al.\ 2009, 2010a, 2010b; Yoon et al.\ 2011).
 
We start our analyses by assuming two sets of CR proton and electron spectra at the heliopause, one for the Galactic CRs (``Galactic'') and the other for the CRs from local CVs and novae (``local''). The ``local'' represent CRs accumulated in the local bubble from historic local CVs and novae. We ignore CRs that might have been injected by supernovae within the CR trapping time of the local bubble (10$^6$ to 10$^7$ years).  

\begin{figure*}
 \begin{center}
   \includegraphics[width=12cm]{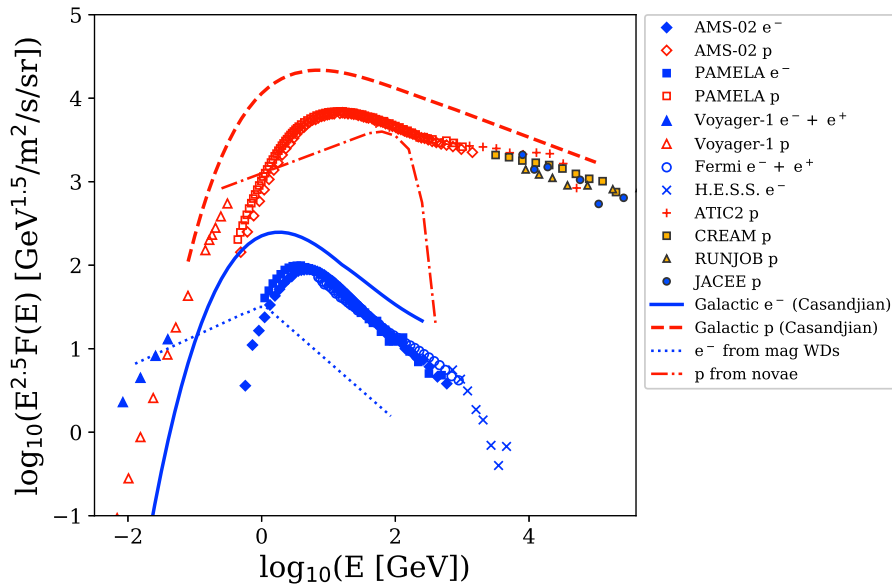}
\end{center}
\caption{The initial Galactic electron and proton spectral shapes deduced from the gamma-ray emissivity in the Galaxy and the initial local spectra deduced from the hard X-ray spectra at the CVs and the gamma-ray spectrum at the nova. Data points are from CR measurements near Earth and at the heliopause. The spectra are depicted by blue solid line for Galactic electron, red dashed line for Galactic proton, blue dotted line for local electron, and red dot-dash line for local proton. Here the CR spectra are not normalized to the data.
}\label{fig1}
\end{figure*}

In our model, the logarithm of the differential CR fluxes are multiplied by the 2.5-th power of CR kinetic energy and represented by polynomial functions of degree 6 or less in log-10 of CR kinetic energy. For energies $>$ 400 GeV, the Galactic proton spectrum is assumed to approach a single power-law (PL) form. The polynomial coefficients for the Galactic spectra are constrained initially by fitting the proton and electron spectra deduced from gamma-ray emissivity in the local Galaxy (the latitude $10^\circ<|b|<70^\circ$) by Casandjian (2015). 
Our initial model spectra for the local electron and proton spectra adopt the following shapes respectively (see Figure~\ref{fig1}): a broken PL for the local electron spectrum implied by the observed synchrotron hard X-ray emission from magnetized WDs (Terada 2013; Ferrario et al. 2015; Geng, Zhang, \& Huang 2016), and a single PL with an exponential cutoff for the local proton spectrum suggested by the observed gamma-ray spectrum of novae (Abdo et al.\ 2010). These spectra are reshaped regressively until the sums of the Galactic and local spectra reproduce the observed CR spectra in the energy range least affected by the Sun (E $>$ 100 GeV) and the Voyager-1 data points taken outside of the solar magnetosphere (Stone et al.\ 2013; Cummings et al.\ 2016). 

Once the proton Galactic and local model spectra are established, we extend them to nuclear CRs and address topics such as the hardening in the nuclear CR spectra (Ahn et al.\ 2009, 2010a, 2010b, Yoon et al.\ 2011). The gamma-ray emissivity added by the CRs from CVs and nova outbursts in the local bubble and the Galaxy are also studied.

The CR spectral shapes have been analyzed extensively in the literature in the past several years. Most studies focused on the positron-electron ratios as summarized in Accardo et al.\ (2014) for experiments and in Cholis and Hooper (2013) and Boudaud et al.\ (2015) for their theoretical interpretations. Although the positron-electron ratio is related to the present study, it will not be discussed in this paper. Some other topics discussed in the literature concern our work. They include the low energy electron spectrum by Voyager-1 analyzed by Jaffe et al.\ (2011), Strong, Orlando, and Jaffe (2011), and Orlando et al.\ (2015), propagation processes (Putze, Maurin, \& Donato 2011; Blasi 2013; Potgieter 2013; Grenier et al.\ 2015; Thoudam et al.\ 2016) and the spectral hardening in nuclear CRs (Ahn et al.\ 2010a; Tomassetti 2015 \& 2016; Ohira, Kawanaka, \& Ioka\ 2016; Khiali, Haino, \& Feng\ 2017). In these works, however, no specific CR source was identified to explain these CR spectral features including the spectral hardening in the nuclear CRs.

The study presented here analyses the CR spectra including those emitted from historic CVs and novae, and accumulated in the local bubble. Because of the paucity of high-precision CR spectral data, we limit our analyses to energies less than a few TeV and reduce free parameters or degrees of polynomials to be fitted to. Our modeling differs from the prior works of similar kind in the following points.
\begin{itemize}
\item Voelk and Berezhko (2013) found a surplus in the low energy spectra. Tomassetti (2016) fitted the proton and helium spectra as a sum of sub-TeV and multi-TeV components. However, neither associated the low energy CR component with local CVs/novae nor extracted the spectra for individual CR species.
\item No prior work fitted the local proton and nuclear CR spectra with one common spectral shape as done in this work.
\item Our model identifies historic local CVs and novae to be a major source of sub-TeV CRs, quantifies their CR fluxes, and allows one to compare the CR fluxes from historic CVs and novae in various locations of the Galaxy through gamma-ray emissivity measurements.
\end{itemize}

We note that the CVs and novae detected in the hard X-ray and gamma-ray bands are not within the local bubble. They are assumed to represent historic CVs and novae occurred in the bubble over the CR trapping time of 10$^6$ to 10$^7$ years (Strong, Moskalenko, \& Ptuskin 2007; Erlykin, Makavariani, \& Wolfendale 2017). Later, we will estimate the energies carried out from CVs and novae as CRs and compare with the total energy stored as the local CRs. We then argue for the validity of the assumptions.

\section{Model spectra for Galactic electron and proton CRs at the heliopause}

The CR spectra for electrons and protons at the heliopause are assumed to be the sum of one Galactic component and one local component for each species. Our knowledge on CR spectra comes from near-Earth observations except for those measured by Voyager-1 at the heliopause. The spectra observed near Earth are known to be distorted along the propagation path from the heliopause to Earth. The distortion is known to depend on CR's energy and charge, and, to change in time. We avoid this complication by modeling CR spectra at the heliopause. 

We make a few other simplifications to minimize degeneracy between the Galactic and local spectra. One simplification is that the log-10 of Galactic CR spectra multiplied by the 2.5-th power of CR kinetic energy are constrained to polynomials with degree of 4 of log$_{10}$(E [GeV]). This constraint facilitates decoupling between the Galactic and local spectra as will be explained below. 
\begin{enumerate} 
  \item We start with the Galactic spectra for electrons and protons deduced from the diffuse gamma-ray emissivity in the local Galaxy (the latitude $10^\circ<|b|<70^\circ$) given by Casandjian (2015). The author derived the local interstellar spectra (LIS) combining the emissivity in the local Galaxy, the CR spectra taken near Earth, the solar modulation effect, and the propagation effect from the local Galaxy. The force-field approximation (Gleeson \& Axford 1968) was used to demodulate the solar modulation effect and GALPROP (Strong \& Moskalenko 1998a, 1998b) to estimate the propagation effect. The spectral shapes are shown in Figure 1 by the blue solid line (electron) and the red dashed line (proton). These are not normalized to the data.
  \item The CR spectra described above were not constrained by the Voyager-1 data (Stone et al.\ 2013; Cummings et al.\ 2016) taken at the heliopause nor by the high-energy CR spectra not affected by the solar modulation.  We will normalize our model spectra to the high energy CR data (E $>$ 100 GeV) and extend downward in the flux to meet the Voyager-1 data points in the sub-GeV to GeV range later in the fitting process.
  \item Our initial spectrum for the Galactic electron rolls down sharply below 1 GeV (the blue solid line in Figure 1) and requires a large low energy local flux to reach the Voyager-1 data points. This is quite consistent with the finding by Lee et al.\ (2011) that the low energy electron spectrum observed near Earth shows an ``excess'' when compared with the spectra expected from the nominal electron CR sources.
  \item Our initial spectrum for the Galactic proton is significantly higher than the Voyager-1 data point if extended downward in energy (the red dashed line in Figure 1).
  \item For E $>$ 100 GeV, the Galactic electron spectrum will be constrained by the HESS data (Aharonian et al.\ 2008). The Galactic proton spectrum is forced to approach to a PL of index $-2.7$, the most popular choice in the literature (Nakamura et al.\ 2010).
  \item The first trial Galactic spectral shapes shown in Figure 1 will be fitted regressively so that the sums of the Galactic and local spectra will reproduce smoothly the Voyager-1 data and the high energy (E $>$ 100 GeV) CR data.
\end{enumerate}
\section{Model spectra for local electron and proton CRs at the heliopause}
We start our local CR electron spectrum from the one inferred by the synchrotron hard X-ray spectra observed at in AE Aquarii and AR Scorpii (Terada et al.\ 2008; Oruru \& Meintjes 2012; Geng, Zhang, \& Huang 2016). Terada et al.\ (2008) concluded that electrons are accelerated to energies higher than $\sim$10 GeV in AE Aquarii. Geng, Zhang, and Huang (2016) analyzed radiation from AR Scorpii detected by Marsh et al.\ (2016) and predicted that the underlying electron spectrum has a broken PL shape and the difference between the higher and lower PL branches to be $\sim -1$. Oruru and Meintjes (2012) studied the data on AE Aquarii taken with Suzaku, Chandra, and Swift to conclude that electrons are accelerated to $\sim$100 GeV. Based on these results, we adopt a broken PL spectrum with lower and higher PL indices of $-2.13$ and $-3.2$, respectively, with a break at 1 GeV as the initial local electron spectrum as shown by the blue dotted line in Figure 1. We note the breaking energy is set higher than that predicted by Geng, Zhang, and Huang (2016) for AR Scorpii because the electron spectrum was analyzed to reach $\sim$100 GeV in AE Aquarii by Oruru and Meintjes (2012).

We adopt the local proton spectrum extracted from the gamma-ray spectrum observed at V407 Cygnii 2010 (Abdo et al.\ 2010) as the initial trial function. The proton spectral shape was predicted to be of single PL shape with index = $-2.15^{+0.28}_{-0.45}$ and cut off exponentially at $32^{+85}_{-8}$ GeV. We set the PL index to $-2.2$ and the exponential cut-off at 150 GeV in the initial trial (the red dot-dash line in Figure 1). The later discoveries of multi-GeV gamma-rays from other novae (Ackermann et al.\ 2014; Cheung et al.\ 2016) are consistent with our initial spectrum.

The logarithm of the initial local spectra multiplied by the 2.5-th power of CR kinetic energy are fitted by polynomials of degree 6 with log-10 of kinetic energy as the variable, in the fitting process described in the next sub-section.

\section{Fitting the sum of the Galactic and local CR spectra to the observed data}

The Galactic and local CR spectra overlap in the middle energy range. The degeneracy between the two components is quite significant for the proton spectrum.  The reduction of the degree of polynomials to 4 for the Galactic electron and proton spectra described before facilitates the fitting process. The Galactic spectra are first fitted in the high energy portion ($E>100$ GeV). The lower energy portion of the Galactic spectra, however, overlaps with the respective local spectra which requires careful fitting to the data in the high energy band ($E>100$ GeV) and the Voyager-1 data after summing the two components.

The Galactic electron spectrum is reshaped at higher energies to match the the higher energy ($E>100$ GeV) electron data including the HESS data (Aharonian et al.\ 2008). At lower energies, the initial Galactic electron spectrum by Casandjian (2015) was substantially lower than that of Voyager-1. The lower-energy branch of the initial broken PL electron spectrum inferred by the hard X-ray spectra from the CVs (the blue dotted line in Figure 1) gave values higher than the Voyager-1 data. It is normalized and reshaped to pass the Voyager-1 data in the polynomial shape. The higher-energy branch of the broken PL local electron spectrum was, however, left unconstrained.  The unconstrained higher-energy branch of the local electron and the unconstrained lower-energy Galactic electron spectra are determined so that their sum connects between the higher energy ($E>100$ GeV) electron data and the Voyager-1 data as a sum of the polynomial forms. 

The total electron spectrum thus obtained is shown by the solid blue line with `+' marks in Figure 2 and listed in a polynomial form in Table 1. The corresponding Galactic and local electron spectra are shown by the blue dashed and blue dotted lines respectively. The formulae are listed in Table 1. The summed spectrum turns out to be stable because the two unconstrained branches, the lower-energy branch of the Galactic spectrum and the higher-energy branch of the local spectrum have very different dependencies on log-10 of energy.

The degeneracy between the Galactic and local proton spectra is harder to decouple. The Galactic proton spectrum is constrained well by ($E>100$ GeV) CR data at higher energies. The initial spectrum taken from Casandjian (2015) gave differential flux values much higher than the Voyager-1 data. The local proton spectrum is predicted to reach its maximum around 150 GeV. It has to roll-down sharply below 1 GeV because the pion production cross-section decreases to zero around log-10(E) of -0.3. The assumption based on the nova observation (Abdo et al.\ 2010) that the local proton spectrum a single PL type of index $-2.2$ and cut-off at around 150 GeV becomes important in constraining the local proton spectrum in a polynomial form. Here again the assumption that the Galactic proton spectrum is represented by a polynomial of degree 4 has helped to reduce the degeneracy between the Galactic and local spectra.

The total proton spectrum is shown by the solid red line with `x' marks in Figure 2 and listed in polynomial forms in Table 1. The corresponding Galactic and local spectra are shown by the red long-dash and red dot-dash lines respectively. Their formulae are given in Table 1. 

Below summarizes our observations from the CR spectra obtained in our polynomial regression: 
\begin{enumerate} 

\item The electron CR spectrum measured by Voyager-1 is almost all accounted by the electrons injected from CVs and accumulated in the local bubble. This gives an answer to the question raised by Lee et al.\ (2011) that the low energy electron spectrum observed near Earth shows an ``excess" relative to the standard electron spectrum commonly used, for example, in Accardo et al.\ (2014), Cholis and Hooper (2013), and Boudaud et al.\ (2015)

\item If the low energy electron flux measured by Voyager-1 had been associated with the Galactic component, the high flux would have conflicted with the observed synchrotron radio emission as discussed by Webber and Higbie (2008). 

\item Our final local electron spectrum agrees with the broken PL spectral shape predicted by Geng, Zhang, and Huang (2016) except that the breaking energy has turned out to be higher. The breaking energy must be high to be consistent with the analysis by Oruru and Meintjes (2012) that the electron spectrum reaches $\sim$100 GeV in AE Aquarii.  

\item The proton spectrum measured by Voyager-1 is dominated by the Galactic component because the Galactic CR spectrum is substantially softer (PL index $\sim -2.7$) than the harder spectrum inferred from the observed gamma-ray spectrum at the nova (PL index $\sim -2.2$).

\item Our final local proton spectrum is harder than $-2.7$ between a few 100 MeV and $\sim$10~GeV and consistent with the spectrum inferred from the gamma-ray spectrum at V407 Cygnii 2010 (the red dot-dash line Figure 2).

\item At higher energies ($E> 100$ GeV), almost all CRs are of Galactic origin.

\end{enumerate}

\begin{figure}
 \begin{center}
   \includegraphics[width=8cm]{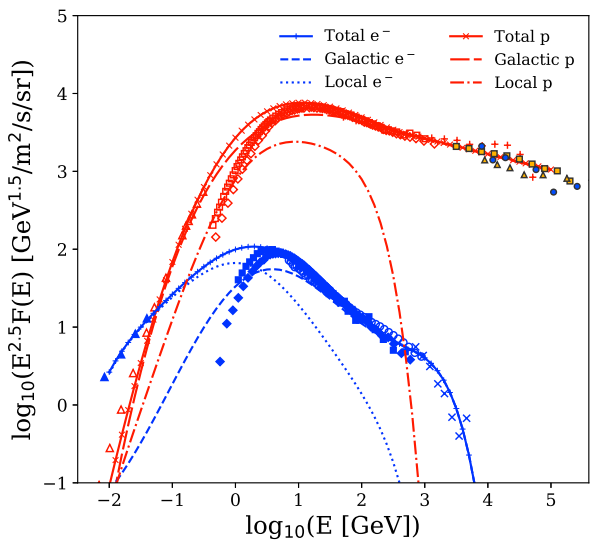} 
 \end{center}
\caption{Model spectra fitted to the CR data in the energy range of E $>100$ GeV and Voyager-1 data in polynomial forms (see Table 1). The Galactic and local CR spectral components are shown separately for both electrons (blue dashed and dotted lines) and protons (red dashed and dash-dotted lines), and the respective total spectra are shown by solid lines. The data points are the same as in Figure~\ref{fig1}.
}\label{fig2}
\end{figure}

\subsection{Extension of our modeling to the nuclear CR spectra at the heliopause}  
Encouraged by the success in reproducing the proton CR spectrum at the heliopause as a sum of the Galactic and local components, we extend our modeling to nuclear CRs. The He, C, N, and O spectra were measured by Voyager-1 at the heliopause (Cummings et al.\ 2016) and by AMS02 recently (Yan et al.\ 2017). Together with other observational data (PAMELA: Adriani et al.\ 2011b; AMS-02: Aguilar et al.\ 2015b, Aguilar et al. \ 2017, ATIC-2: Wefel et al.\ 2008; CREAM: Ahn et al.\ 2009, 2010a, 2010b; Yoon et al.\ 2011), we build spectral models for nuclear CR spectra from sub-GeV to a few TeV per nucleon. We note that the chemical composition of WDs is known to be rich in helium, carbon and heavier nuclei. Our modeling including the local CRs emitted by novae and accumulated in the local bubble is most interesting. 

Since there is no information about nuclear CR spectra in novae, we start with a simplest assumption that all local nuclear CRs have the same spectral shape as the proton local CR. For the Galactic nuclear components, we use the Galactic proton spectral shape as the initial formula for the polynomial regression. The polynomial fitting proceeds straight-forwardly and the PL index comes out to be $\sim -2.58$ for He and $\sim -2.51$ for C, N, and O. For the latter, we choose, for simplicity, a common value of $-2.51$, a value consistent with the ATIC-2 data (Ahn et al. 2009, 2010a) and AMS02 data (Yan et al.\ 2017, Aguilar et al.\ 2017). The fitting has proceeded straight-forwardly and the final model spectra are shown in Figure 3 and tabulated in Table 1. We emphasize that the Galactic and local spectral shapes are assumed to be common to C, N and O. The former differs only slightly from the Galactic proton and He spectral shapes and the latter is constrained to the same shape as the local proton and He CRs. The fact that these similar spectra reproduce all nuclear CR spectra suggests a common origin and acceleration mechanism in novae as will be discussed later in this paper.

\begin{figure*}
\begin{center}
   \includegraphics[width=12cm]{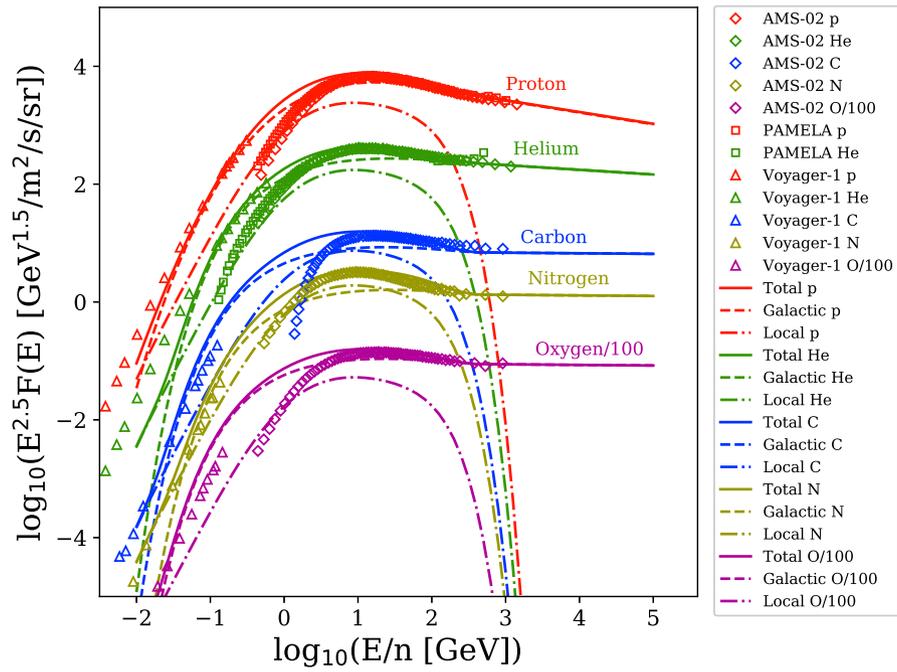} 
 \end{center}
\caption{The total, Galactic and local nuclear CR model spectra are compared with observed data. The Galactic (dashed) and local (dash-dotted) components are shown separately for each CR species, and their respective sums are shown by the solid lines. 
}\label{fig3}
\end{figure*}

\section{Number fluxes for the Galactic and local CRs at the heliopause}

The number fluxes for Galactic and local CRs are calculated for all species on the formulae given in Table 1.  Figure 4 shows the differential number flux per log-10 of kinetic energy in GeV per nucleon. These number spectra are integrated from 10 TeV/n downward to give the CR number fluxes with energy higher than the log-10 of kinetic energy given on the x-axis. The fluxes integrated between 10 MeV/n and 10 TeV/n are given in Table 2. Their ratios relative to the proton fluxes are compared with the solar abundances given in Table 1 of Anders and Grevesse (1989) and the abundances of nova ejecta predicted by Li et al.\ (2016). We note that the integrated fluxes in Table 2 for the Galactic and local CRs do not add-up exactly to the total fluxes because of numerical errors in the polynomial fittings and integrations.

\begin{figure*}
\begin{center}
   \includegraphics[width=12cm]{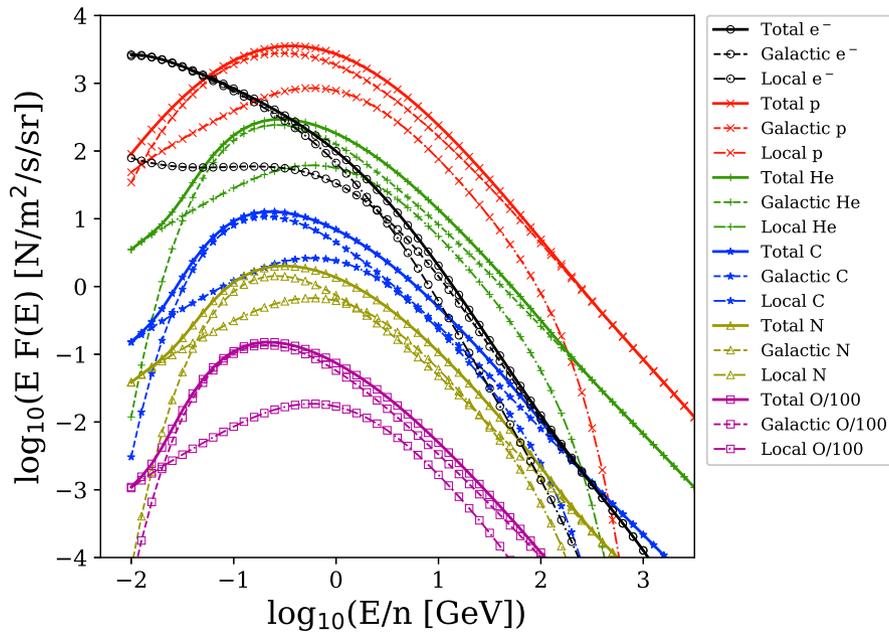} 
\end{center}
\caption{
Comparison of our model spectra for different CR species in the form of differential number fluxes, separated into the respective Galactic and local components. These fluxes are integrated between 10 MeV/n and 10 TeV/n and tabulated in Table 2.
}\label{fig4}
\end{figure*}

\subsection{Energy injected from CVs/novae and energy carried by the local CRs}

The magnetic WDs and gamma-ray novae we used to build our model lie on the boundary or outside of the local bubble. A question remains whether the local electron and proton CR spectra listed in Table 1 are consistent with predictions on the observed CVs and novae. Specifically, we compare the energies likely to have been injected from historic CVs and novae to the local bubble with the energies carried by the local electron and proton CR spectra given in Table 1.

Terada et al.\ (2008) provided an estimate for the luminosity of the synchrotron emission at AE Aquarii in the band of 4 to 30 keV to be $(5-15) \times 10^{30}$~erg/s, assuming a distance of 102 pc. After extending the spectral range to 100 GeV, Oruru and Meintjes (2012) estimate the total energy emitted by the CV star to be 9 $\times$ 10$^{30}$~erg/s. The synchrotron efficiency or the fraction of the electron energy emitted as synchrotron radiation was estimated to be 10$^{-5}$ to 10$^{-4}$ for neutron-star pulsars observed by Chandra (Table 3 of Kargaltsev et al.\ 2012). If the pair multiplication does not develop in the rotating magnetic WD, about 10$^{35}$ to 10$^{36}$ erg/s are likely to have been carried out by CR electrons. If this energy outflow continues for 10$^{4}$ yrs in the local bubble for the electron CR lifetime of $\sim$10$^{6}$ years, a total energy of 3 $\times$ 10$^{46}$ to 3 $\times$ 10$^{47}$ erg or 1.9 $\times$ 10$^{58}$ to 1.9 $\times$ 10$^{59}$ eV will accumulate as the trapped electron CR flux. We assume that the magnetic WDs in the local bubble are represented by AE Aquarii (102~pc from the Sun) and AR Scorpii (116~pc from the Sun). We note that the 16 out of the 17 magnetic WDs detected in the hard X-ray band (Revnivtsev et al.\ 2008) are located at 166~pc to $\sim$2~kpc from the Sun.

Shafter (2017) estimated the Galactic nova rate down to magnitude less than 2 to cover the entire Galaxy. Assuming that the property of Galactic novae does not change over the Galaxy, the nova rate is predicted to be 50 to 100 or more per year. Morris et al.\ (2017) showed that the novae detectability in gamma-ray is limited to a few kpc by the gamma-ray background and considered it not surprising that only 6 were detected by Fermi-LAT while 69 novae were detected in the optical band in the same period. On this, we assume all optically-detected novae eject a flux of CRs comparable to V407 Cygnii 2010. The nova rate estimated for the Galactic disk by Shafter (2017) is $\sim$50 to 500 novae within 50 pc of the Sun in 10$^6$ to 10$^7$ years. Since the lifetime of 10 GeV electron and proton CRs are calculated to be 10$^{6}$ to 10$^{7}$ years (Webber 2015), we expect the CR flux accumulated in the local bubble to be 50 to 500 times the CRs ejected from the nova V407 Cygnii 2010.

Abdo et al.\ (2010) estimated that the energy carried by gamma-rays from V407 Cygni is 3.6 $\times$ 10$^{41}$ erg and by CR protons more than 2.0 $\times$ 10$^{43}$ erg. A significant fraction of CR protons could have been ejected without interacting with gas in the environment. If such a nova occurs every 10$^4$ years in the local bubble as estimated by Shafter (2017) and Morris et al.\ (2017), CR protons will carry out 6.7 $\times$ 10$^{31}$~erg/s from nova outbursts on average. This means 2.0 $\times$ 10$^{45}$ to 2.0 $\times$ 10$^{46}$ erg or 1.2 $\times$ 10$^{57}$ to 1.2 $\times$ 10$^{58}$ eV accumulates during the local proton CR lifetime of 10$^{6}$ to 10$^{7}$ yrs. Classical and recurrent novae observed prior to $\sim$2010 are shown in Figure 1 of Imamura and Tanabe (2012).

The integrated CR number fluxes in Table 2 predict the energy densities of electron and proton CRs at the heliopause to be $\sim$10$^{-3}$ eV/cc and $\sim$10$^{-2}$ eV/cc, respectively. The total energy carried by the CRs in the local bubble (assumed radius $\sim$46 pc and assumed volume of $\sim$1.1 $\times$ 10$^{61}$ cm$^{3}$) is then $\sim$1.1 $\times$ 10$^{58}$~eV for CR electrons and $\sim$1.1 $\times$ 10$^{59}$~eV for CR protons. These energies are not inconsistent for electron CR but about 8 times higher for proton CR when compared with the estimations given in the previous paragraphs for the injections from local magnetic WDs and local historic novae.

We acknowledge that uncertainties remain in the estimations given in this section. It is quite likely, as Sokoloski et al.\ (2017) pointed out, that many more symbiotic stars and novae dim in the optical band may be accelerating particles. Our estimations will need to be revised as new data become available.

\section{Results from CR spectral modeling}

The CR model spectra we obtained (Table 1) bring new knowledge and suggest interesting interpretations as itemized below:

\begin{enumerate}
\item The integrated local electron CR flux is approximately the same as that of the local proton CR flux. Such is not generally expected at the diffusive shock acceleration (DSA) sites like supernova remnants. One interpretation is that magnetized rotating WDs (CVs) are embedded with some mechanism (e.g., magnetic induction) to accelerate electrons quickly to multi-MeV past the Coulomb barrier.

\item The local proton CR spectral shape is applied satisfactorily to all local nuclear CRs. This suggests a common cooling mechanism is operating at the proton/nuclear accelerators in the gamma-ray novae (Vurm \& Metzger 2016).

\item The ``hardening'' of the proton and nuclear CR spectra is largely due to the cut-off setting in at around 100 -- 200 GeV/n in the local proton and nuclear CR spectra.

\item The flux ratios of C, N, and O to proton are substantially higher than those of the solar abundance but much lower than those predicted for novae by Li et al.\ (2016), both for the Galactic and local CRs. This may suggest that the CR accelerators operating in novae are fed not only with gas near the WD surface but also with interstellar matter (ISM) and dust whose abundances are close to the solar as predicted by Ellison, Drury, and Meyer (1997). One likely possibility is that the historic CVs and novae in the local bubble is supplying the excess of proton and nuclear CRs in the energy range $<$ 200 -- 400~GeV and the nuclear CR sources discussed by Biermann and Cassinelli (1993), Ellison, Drury, and Meyer (1997), Parizot et al.\ (2004), Grenier et al.\ (2015), and Ohira, Kawanaka, and Ioka (2016) in the energy range $>$ 200 -- 400~GeV.
\end{enumerate}

\subsection{Comparison with previous works on the CR electron and proton spectra}

Our total electron CR spectrum is compared in Figure 5 with the LIS models given by Bisschoff and Potgieter (2014) and Ackermann et al.\ (2015). Models from the former group resemble more with ours because both include the Voyager-1 data in the fitting process. The difference is in the interpretation of the spectral transition below $\sim$100 MeV. In their interpretation, it is due to Galactic propagation effects while in ours it is due to the local historic CVs and novae contributions. The LIS models for electron given by Ackermann et al.\ (2015) differ much from ours mostly because they did not include Voyager-1 data in the analysis.
\begin{figure}
\begin{center}
   \includegraphics[width=8cm]{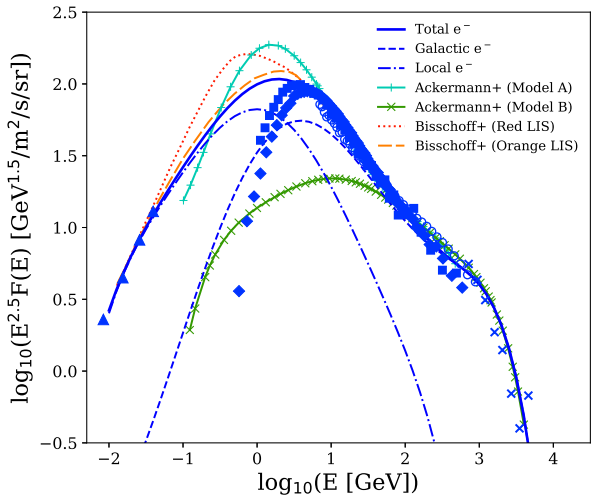} 
\end{center}
\caption{The total electron CR model spectrum obtained in this work (Figure 2 and Table 1) is compared with those given in literature. The orange dashed line and red dotted line are the LIS models by Bisschoff and Potgieter (2014) (see their Table 1 for definition). The cyan solid line with `+' marks and green solid line with `x' marks show the Model A and B in Ackermann et al.\ (2015) respectively. The data points are the same as in Figure~\ref{fig1}.
}\label{fig5}
\end{figure}

Our proton spectrum is compared with those obtained in prior works in Figure 6. Several authors have tried to deduce the LIS from the spectra of proton and/or nuclear CRs near Earth. These works can be classified in three categories: the first type uses observed CR spectra near Earth and corrects for the solar modulation (e.g., Shikaze et al.\ 2007). They did not use the Voyager-1 data to constrain the LIS. The LIS by Shikaze et al.\ (2007) is shown by the blue short dash line in Figure 6.

The second type analyses on the proton CR assume that the spectrum observed near Earth is a sum of a low-energy component and a Galactic component, similarly to the model presented here (e.g., Tomassetti 2016). The Galactic and low energy CR spectra by Tomassetti (2016) are shown in Figure 6 by the green solid line with `+' marks (total) and the green dot-dash line with asterisks (local). The author did not associate the local CRs with the contribution of the local historic novae.

Analyses of the third type deduced the proton LIS from gamma-ray observations relying on GALPROP for source distributions, injection spectra, propagation in the Galaxy (e.g., Johannesson et al.\ 2016). Their LIS is shown by the purple long dash line in Figure 6. The Voyager-1 data were not included in the analysis.

When we compare our work with those discussed above, the enhancement below $\sim$50 GeV in the CR proton spectrum is less pronounced in ours. This is probably because our model spectrum includes the local proton CR with a hard PL index, $-2.2$, below $\sim$20 GeV. We note that all LIS spectra deduced from near-Earth CR observations assumed a softer PL spectrum with an index $\sim$2.75 in correcting for the solar modulation.

Since our model spectra are defined at the heliopause, just outside of the solar system, they can be compared with future near-Earth observations by applying the solar modulation effect applicable at the time of the observation.
\begin{figure}
\begin{center}
   \includegraphics[width=8cm]{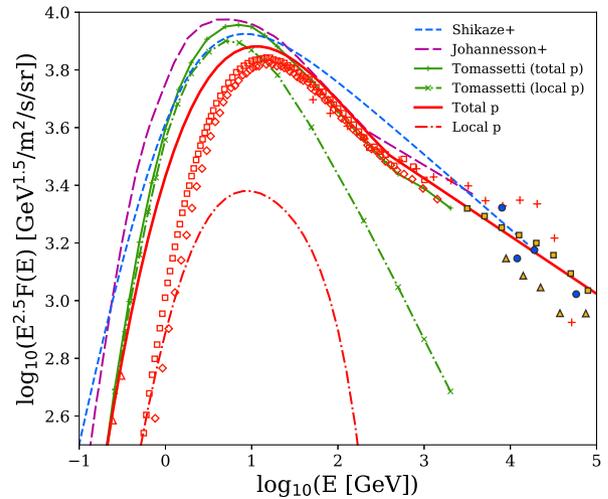} 
\end{center}
\caption{Our model CR proton spectra (total spectrum shown by the red solid line and the local component by the red dash-dotted line), compared with the LIS given by prior studies. 
See text for detailed description of the lines. The data points are the same as in Figure~\ref{fig1}.}
\label{fig6}
\end{figure}

\section{Diffuse gamma-ray spectra produced by CRs from CVs and novae}

The CRs from CVs and novae interact with ISM gas and produce diffuse gamma-rays. For us located near the center of the local bubble, they are distributed nearly isotropic. Our local CR spectra will be tested against the isotropic gamma-ray spectrum extracted from Fermi-LAT data (Ackermann et al.\ 2015). Such gamma-rays are also expected in other locations of the Galaxy if CRs from CVs and novae are accumulated there. One location is the inner Galaxy where the ISM gas density and WD density is likely to be higher than in the local bubble..

\subsection{Gamma-ray spectrum produced by the local CRs from CVs and novae}

The spectrum of diffuse gamma-rays, distributed isotropic around the solar system is calculated, assuming that a sphere of radius 46 pc is filled with local electron and proton CRs with the spectra given in Table 1. The gas density is assumed to be 0.05 H per cm$^3$. Gamma rays are assumed to be produced by proton-proton via decays of neutral pions with the cross-section given by Kamae et al.\ (2006) and by electron-proton bremsstrahlung with the cross-section given by Koch and Motz (1959). The contributions of nuclear CRs and target nuclei are taken care by multiplying the gamma-ray spectra obtained for the proton-proton and electron-proton interactions with the respective nuclear factors calculated for the composition of the CRs from CVs and novae in Table 2. The predicted gamma-ray spectrum is compared with the isotropic gamma-ray background (Ackermann et al.\ 2015) extracted from Fermi-LAT observations.

\begin{figure}
\begin{center}
   \includegraphics[width=8cm]{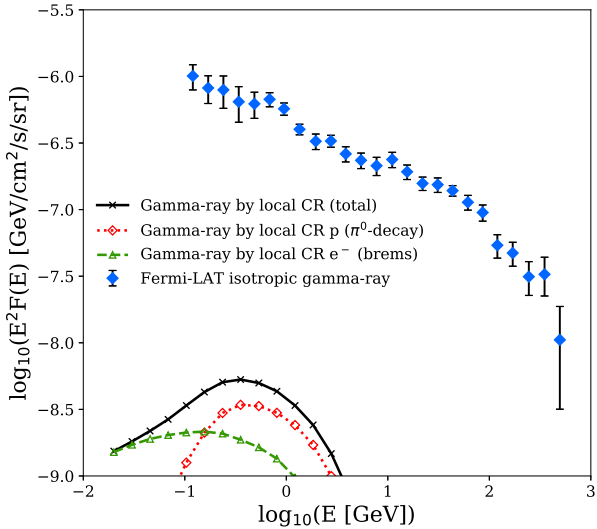} 
 \end{center}
\caption{The isotropic gamma-ray spectrum extracted from Fermi-LAT observations (Ackermann et al.\ 2015, shown by blue diamonds) and that expected from the local electron, proton, and nuclear CRs interacting with ISM in the local bubble of radius 46 pc and density 0.05 H/cm$^3$. The black solid line shows the total contribution of the local CRs to the isotropic gamma-ray spectrum, which is composed of a $\pi^0$-decay component from the local protons and nuclei (red dotted line), and a bremsstrahlung component from the local electrons (green dashed line).}\label{fig7}
\end{figure}

Shown in Figure 7 are the isotropic gamma-ray background spectrum (blue diamonds) from Fermi-LAT observations (Ackermann et al.\ 2015). The total gamma-ray spectrum expected for local CRs (listed in Table 1), its neutral pion component, and its bremsstrahlung component are also shown in the figure. The neutral pion component is calculated for the proton-proton interaction and multiplied with a nuclear factor (1.92) for combination of the CR composition given in Table 1 and the ISM composition of Meyer (1985).  Note that the factor is higher than 1.84 suggested by Mori (2009) because of the higher mix of C, N, and O in the local CR. The factor for the bremsstrahlung component is 1.5 which accounts for the nuclei in the ISM.

We find that the total gamma-ray spectrum expected for the local CR contribution is too low to be visible in the diffuse isotropic gamma-ray spectrum. The expected intensity is low because the ISM density in the local bubble is roughly 10 to 20 times lower than that in the local or inner Galaxy disk, respectively (Nakanishi \& Sofue 2003, 2006). The spectral shape, however, is quite distinct and shows a hump-like shape between sub-GeV and a few GeV (the ``GeV hump"). In high density gas environment, the gamma-rays generated by CRs from CVs and novae are likely visible as discussed in the next sub-section.

\subsection{``GeV hump" in the diffuse gamma-ray spectrum from the inner Galaxy}
Recent observations in the hard X-ray band strongly suggest that WDs are contributing to the Galactic ridge hard X-ray emission (Yuasa, Makishima, \& Nakazawa 2012; Revnivtsev et al.\ 2008; Krivonos, et al.\ 2012). Motivated by these findings, we search for the ``GeV hump," characteristic of the local CRs shown in the bottom-left corner in Figure 7. For this we compare the diffuse gamma-ray emissivity obtained in different regions of the Galaxy.

To quantify the difference in CR spectra by comparing diffuse gamma-ray spectra in two regions, the spectra must be normalized to the amount of target material. This is done by dividing the gamma-ray spectrum by the number of hydrogen atoms (atomic HI + 2 $\times$ molecular H$_2$) in the region, which is called gamma-ray emissivity per hydrogen atom. Figure 8 shows the observed gamma-ray emissivity in the region of Galactocentric radius (R) $<$ 1.5 kpc by red diamonds (taken from Figure 7b of Acero et al.\ 2016). The solid black line is their model based on GALPROP (Strong \& Moskalenko 1998a, 1998b) for the region and the blue dashed line is that for R = 8 -- 10 kpc. The red dotted line represents the difference between the two emissivity spectra for R $<$ 1.5 kpc and R = 8 -- 10 kpc. 

\begin{figure}
\begin{center}
   \includegraphics[width=8cm]{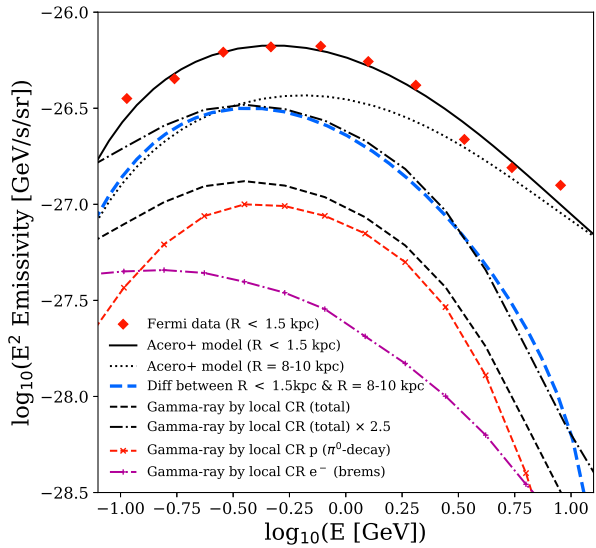} 
 \end{center}
\caption{Gamma-ray emissivities per H-atom for Galactocentric radii R $<$ 1.5 kpc and R = 8 -- 10 kpc (Acero et al.\ 2016), and the difference between the two. The Fermi-LAT data for R $<$ 1.5 kpc are shown by red diamonds. Model fits for the two Galactocentric rings are shown by the black solid (R $<$ 1.5 kpc) and black dotted (R = 8 -- 10 kpc) lines. The blue dashed line represents the difference between the two emissivity spectra. The total emissivity estimated for our local CR model is represented by the black dashed line, which is again decomposed into a $\pi^0$-decay component by CR protons and nuclei (red dashed with `x') and a bremsstrahlung component by CR electrons (purple dot-dash with `+'). The total emissivity curve for the local CRs is then multiplied by a factor of 2.5 (black dash-dotted line), which can satisfactorily account for the apparently enhanced GeV emissivity in the inner Galaxy (i.e., blue dashed line) relative to the outer region. 
}
\label{fig8}
\end{figure}

Acero et al.\ (2016) calculated the emissivity for the neutral pion component following the procedure described in Mori (2009) and Casandjian (2015): the flux ratios of major nuclear CR species relative to proton (He, CNO, NeMgSiS, and Fe) are taken from Honda et al.\ (2004) at 10 GeV/n, the ISM composition from Meyer (1985), and the nuclear multiplication factors for the projectile-target combinations from Mori (2009). For the bremsstrahlung component, the nuclear factor is different from that for the neutral pion component. Acero et al.\ (2016) used the formula in GALPROP to account for the difference. Since our local CR composition differs from the nominal values (in this case that used in Acero et al.\ 2016), conversion from the gamma-ray spectrum to the gamma-ray emissivity needs to be recalculated.

The gamma-ray spectra (fluxes) for the local CRs shown in Figure 7 are converted to the emissivity assuming the CR abundances given in the right-most 3 columns of Table 2 for protons, He, and CNO, and those given by Honda et al.\ (2004) for NeMgSiS and Fe at 10 GeV/n. The ISM composition of Meyer (1985) and the nuclear multiplication factors for the projectile-target combinations given by Mori (2009) are adopted. The results are shown in Figure 8 by the black dashed line (total), the red dashed line with symbols `x' (neutral pion), and the purple dot-dash line with symbols `+' (bremsstrahlung). The thick black dot-dash is the emissivity for the WD CR fluxes 2.5 times the local CR fluxes given in Table 1. The line agree with the difference between the inner Galaxy and the local Galactocentric ring in the energy range E $>$ 0.3 GeV. This strongly suggests that CR fluxes from WDs are higher in the inner Galaxy than in the ring R = 8 -- 10 kpc, in general agreement with the hard X-ray observations by Yuasa, Makishima, and Nakazawa (2012) who fitted the Galactic Ridge X-ray Emission with a spectral model of magnetic accreting WDs, or CVs, with a mass of 0.66$^{+0.09}_{-0.07}$ M$_{\odot}$. The authors concluded that the Galactic Ridge X-ray Emission is essentially an assembly of numerous discrete faint X-ray sources.

We note that the gamma-ray emissivity spectrum predicted by our local CRs (black dash-dotted line in Figure 8) does not agree with the emissivity difference between the two regions of the Galaxy for E $<$ 0.3 GeV. This may be due to the difference in the electron-to-proton ratio in the two regions.

\section{Conclusions}

We have constructed a set of spectral models for the CRs at heliopause for electrons and protons. The CR spectra are assumed to consist of the Galactic spectra and the local spectra. They are fitted in polynomial regression starting with the initial Galactic spectra deduced from the Galactic gamma-ray emissivity observed by Fermi-LAT (Casandjian 2015) and the initial local spectra from the hard X-ray spectra at two CVs (Terada et al.\ 2008, Marsh et al.\ 2016) and the gamma-ray spectrum at a nova (Abdo et al.\ 2010).  By limiting the Galactic electron and proton spectral shapes to polynomials of degree 4 and the local shapes to polynomials of degree 6, we could determine the two components by constraining their sum to reproduce the high energy spectral data (E $>$ 100 GeV) and the Voyager-1 data, species by species.

The CR spectral modeling is then extended to nuclear CRs (He, C, N, and O) under a simplest assumption that the local He, C, N, and O spectra have the same shape as the local proton spectrum and that the Galactic C, N, and O spectra have another common shape very similar to those of Galactic proton and He.

The final model spectra are shown in Figure 2 and 3, and given in polynomial formulae in Table 1. The following conclusions are drawn from our study:
\begin{enumerate}
\item The ``hardening'' in the proton and nuclear CR spectra is due to a sharp cut-off of the local components at around 100 -- 200 GeV/n. This may suggest existence of a common cooling mechanism (Vurm \& Metzger\ 2016) for protons and nuclei in novae.

\item We find that the interactions of the CRs from CVs and novae with ISM produce a hump-like shape between sub-GeV and a few GeV (the ``GeV hump") in the gamma-ray spectrum as shown in Figure 7. Such a hump is seen in the diffuse gamma-ray emissivity of the inner Galaxy (R $<$ 1.5 kpc). We interpret this as due to high CR fluxes injected from CVs and novae in the region.

\item The local electron CR flux is approximately the same as the proton local CR flux if integrated over the spectra. The electron-to-proton ratio is about two orders of magnitude higher than that commonly assumed for the supernova remnants. One possible explanation is that magnetic induction accelerates electrons quickly to overcome the Coulomb barrier near the surface of rotating magnetic WDs (CVs).

\item The energy densities of the local electron and proton CRs are calculated on our model spectra to be $\sim$10$^{-3}$ eV/cc and $\sim$10$^{-2}$ eV/cc, respectively. These energy densities are small when compared with a crude estimate on the total energy carried by the Galactic CRs ($\sim$1 eV/cc) in the Galactic disk.

\end{enumerate}

The present study leaves some uncertainty in the following two areas. First, the population of magnetic WDs (or CVs) and historical novae are not well known even within 100 pc of the Sun. The second is that CR spectra are poorly known in the multi-TeV range. This situation will be improved as new optical surveys extend their sky coverage and find many more WDs, and, direct CR measurements reach to multi-TeV range.

\begin{ack}
TK gratefully thanks Dr. A. W. Strong for valuable discussion through this work. TK also appreciate Dr. Don Ellison reading a draft of this paper in early stage and offering comments. TK and SHL acknowledges many years of fruitful discussions with the Fermi-LAT team. The authors thank Dr. Y. Ohira for pointing out a numerical error in CR energy calculation in section 5.1. TK has been supported by JSPS Grant (15K05098) during this work.
\end{ack}

\begin{center}
\begin{longtable}{ll}
  \caption{Polynomial representations of our CR spectral models}
  \hline            
  Category & Fitting polynomial \footnotemark[1]  \\  
  \hline 
  \endfirsthead
  \hline
  \endhead
  \hline
  \endfoot
  \multicolumn{2}{l}{$^1$\footnotesize Here x = log$_{10}$(E/n [GeV]) and y = log$_{10}$(E$^{2.5}$ F [GeV$^{1.5}$/m$^2$/s/sr])}\\
  \multicolumn{2}{l}{$^2$\footnotesize See Casandjian (2015)}\\
  \multicolumn{2}{l}{$^3$\footnotesize Power-law with index -2.8 for E/n $>$ 1 TeV}\\
  \multicolumn{2}{l}{$^4$\footnotesize Power-law with index -2.7 for E/n $>$ 400 GeV}\\
   \multicolumn{2}{l}{$^5$\footnotesize Power-law with index -2.58 for E/n $>$ 400 GeV}\\
  \multicolumn{2}{l}{$^6$\footnotesize Power-law with index -2.51 for E/n $>$ 400 GeV}\\
  \endlastfoot
 \multicolumn{2}{l}{\textbf{CR spectra in Figure 1}}\\
 Galactic electrons (Casandjian) \footnotemark[2] 	& $-0.0065x^4 + 0.1895x^3 - 0.726x^2 + 0.3467x + 2.3469\ \ \ (x < 2.4)$ \\ 
 Galactic protons (Casandjian) \footnotemark[2, 3] 	& $0.002x^5 - 0.033x^4 + 0.2143x^3 - 0.6834x^2 + 0.7728x + 4.0566\ \ \ (x > -1.1)$  \\ 
 Electrons from mag WDs  			& Broken PL with indices $-2.13$ $(x < 0)$ and $-3.2$ $(x \ge 0)$\ \ \   $(y = 1.5$ at $x = 0)$ \\     
 Protons from novae 					& PL with index $-2.2$ and exponential cut off at $x = 2$ \\
 \multicolumn{2}{l}{\textbf{CR spectra in Figure 2 \& 3}}\\  
 Total electrons  				& $y = -0.005x^6 + 0.014x^5 + 0.0496x^4 - 0.0783x^3 - 0.4279x^2 + 0.2574x + 1.9971$  \\
 Galactic electrons  				& $y = -0.0042x^6 + 0.005x^5 + 0.0798x^4 - 0.0765x^3 - 0.6399x^2 + 0.7629x + 1.5227$ \\ 
 Local electrons  				& $y = -0.0101x^6 + 0.011x^5 + 0.0844x^4 - 0.0599x^3 - 0.5639x^2 - 0.0015x + 1.8238$ \\ 
 Total protons \footnotemark[4] 		& $y = 0.00002x^6 - 0.00035x^5 + 0.0145x^4 + 0.0714x^3 - 0.5877x^2 + 0.9545x + 3.4315$  \\
 Galactic protons \footnotemark[4] 	& $y = -0.0071x^4 + 0.1021x^3 - 0.5177x^2 + 0.8669x + 3.2712$ \\ 
 Local protons  					& $y = -0.123x^6 - 0.0039x^5 + 0.1047x^4 - 0.0106x^3 - 0.7465x^2 + 1.1583x + 2.89$ \\ 
 Total helium \footnotemark[5]		& $y = 0.0009x^6 - 0.0104x^5 + 0.0213x^4 + 0.1363x^3 - 0.6875x^2 + 0.8761x + 2.3212$  \\
 Galactic helium \footnotemark[5] 	& $y = -0.0015x^6 + 0.0177x^5 - 0.0834x^4 + 0.2299x^3 - 0.4817x^2 + 0.6316x + 2.1$  \\
 Local helium					& $y = y($Local protons$) - 1.14$  \\
 Total carbon \footnotemark[6]		& $y = 0.0003x^6 - 0.0037x^5 + 0.0024x^4 + 0.1332x^3 - 0.6074x^2 + 0.7704x + 0.8867$  \\
 Galactic carbon \footnotemark[6] 	& $y = y($Galactic helium$) - 1.505$  \\
 Local carbon 					& $y = y($Local protons$) - 2.51$  \\
 Total nitrogen \footnotemark[6] 	& $y = 0.0007x^6 - 0.0094x^5 + 0.0319x^4 + 0.0937x^3 - 0.6971x^2 + 0.952x + 0.16$  \\
 Galactic nitrogen \footnotemark[6] 	& $y = y($Galactic helium$) - 2.23$  \\
 Local nitrogen	 				& $y = y($Local protons$) - 3.1$  \\
 Total oxygen \footnotemark[6]		& $y = -0.0126x^4 + 0.147x^3 - 0.5642x^2 + 0.7192x + 0.9$  \\
 Galactic oxygen \footnotemark[6] 	& $y = y($Galactic helium$) - 1.4$  \\
 Local oxygen					& $y = y($Local protons$) - 2.66$  \\
   \hline
\end{longtable}
\end{center}

\begin{center}
\begin{longtable}{|c|cccccc|}
  \caption{Number fluxes\footnotemark[1] of CR species and their ratios from Figure~\ref{fig4}}
  \hline            
   CR  & e$^-$ & H & He & C & N & O \\ 
  \hline 
  \endfirsthead
  \hline
  \endhead
  \hline
  \endfoot
  \multicolumn{7}{l}{$^1$\footnotesize Number fluxes [N/m$^2$/s/sr] integrated from 10 MeV/n to 10 TeV/n}\\
  \multicolumn{7}{l}{$^2$\footnotesize Solar abundance by Anders and Grevesse (1989)}\\
  \multicolumn{7}{l}{$^3$\footnotesize Solar abundance by Li et al.\ (2016)}\\
  \multicolumn{7}{l}{$^4$\footnotesize Abundance for Nova V959 Mon by Peretz et al.\ (2016)} \\
  \multicolumn{7}{l}{$^5$\footnotesize Abundance for ring nebulae around massive stars by Esteban et al.\ (2016)} \\
  \endlastfoot
  Total flux 				& $2326$  	& $5454$ 	& $388.0$ 		& $16.07$ 				& $2.52$ 					&  $17.66$ \\
  Ratio 					& - 			& $1.0$ 	& $0.071$ 		& $2.9 \times 10^{-3}$ 		& $4.6 \times 10^{-4}$ 		& $3.2 \times 10^{-3}$ \\
  \hline
  Galactic flux 				& $130.6$  	& $3790$ 	& $282.2$ 		& $12.46$ 				& $1.66$ 					& $15.87$ \\
  Ratio  					& - 			& $1.0$ 	& $0.074$			& $3.3 \times 10^{-3}$ 		& $4.4 \times 10^{-4}$ 		& $4.2 \times 10^{-3}$ \\
  \hline
  Local flux 				& $2197$  	& $1640$ 	& $116.1$ 		& $3.93$  					& $1.01$ 					& $2.79$ \\
  Ratio      					& - 			& $1.0$  	& $0.071$  		& $2.3 \times 10^{-3}$  		& $6.2 \times 10^{-4}$  		& $1.7 \times 10^{-3}$ \\
    \hline
  A \& G \footnotemark[2] 		& -  			& $1.0$ 	& $0.097$ 		& $3.6 \times 10^{-4}$ 		& $1.1 \times 10^{-4}$ 		& $8.5 \times 10^{-4}$ \\
   \hline
  Li+  \footnotemark[3] 		& - 			& $1.0$ 	& $0.125$ 		& $1.8 \times 10^{-4}$ 		& $0.013$ 				& $0.011$ \\
  \hline
  Peretz+ \footnotemark[4]  	& 			& 		& 				& 						& 						& \\
  X-ray  					& -  			& $1.0$ 	& $0.242$ 		& $2.2 \times 10^{-4}$ 		& $6.0 \times 10^{-3}$ 		&  $0.13$ \\
  Optical  					&  - 			& $1.0$ 	&  $0.127$ 		& - 						& $3.4 \times 10^{-3}$ 		&  $1.1 \times 10^{-3}$ \\
    \hline
    Esteban+  \footnotemark[5]  	& - 			& $1.0$ 	& $0.077 - 0.288$ 	& $(2.6 - 160) \times 10^{-4}$ 	& $(3.4 - 55) \times 10^{-5}$ 	& $(9.5 - 81) \times 10^{-5}$  \\
   \hline
\end{longtable}
\end{center}

\end{document}